\documentclass[
reprint,      
aps,
pre,
amsmath,
amssymb,
]{revtex4-2}

\usepackage{graphicx}
\usepackage{hyperref}
\usepackage{mathrsfs}
\usepackage[version=3]{mhchem}
\usepackage{multirow}
\usepackage[abbreviations,xlatin]{foreign}
\usepackage{adjustbox}
\usepackage{pgfplots}
\pgfplotsset{compat=1.17}
\usepackage{siunitx}

\makeatletter
\@ifpackageloaded{txfonts}\@tempswafalse\@tempswatrue
\if@tempswa
  \DeclareFontFamily{U}{txsymbols}{}
  \DeclareFontFamily{U}{txAMSb}{}
  \DeclareSymbolFont{txsymbols}{OMS}{txsy}{m}{n}
  \SetSymbolFont{txsymbols}{bold}{OMS}{txsy}{bx}{n}
  \DeclareFontSubstitution{OMS}{txsy}{m}{n}
  \DeclareSymbolFont{txAMSb}{U}{txsyb}{m}{n}
  \SetSymbolFont{txAMSb}{bold}{U}{txsyb}{bx}{n}
  \DeclareFontSubstitution{U}{txsyb}{m}{n}
  \DeclareMathSymbol{\aleph}{\mathord}{txsymbols}{64}
  \DeclareMathSymbol{\beth}{\mathord}{txAMSb}{105}
  \DeclareMathSymbol{\gimel}{\mathord}{txAMSb}{106}
  \DeclareMathSymbol{\daleth}{\mathord}{txAMSb}{107}
\fi
\makeatother

\def\equationautorefname#1#2\null{Equation#1(#2\null)}

\begin{document}

\title{Distribution of the Radius of Gyration for an ISAW}

\author{Antony Lesage}
\email{antony.lesage@sorbonne-universite.fr}
\affiliation{Sorbonne Universit\'e, CNRS, PHENIX, F-75005 Paris, France}

\author{Jean-Marc Victor}
\email{jean-marc.victor@sorbonne-universite.fr}
\affiliation{Sorbonne Universit\'e, CNRS, LPTMC, F-75005 Paris, France}

\author{Vincent Dahirel}
\email{vincent.dahirel@sorbonne-universite.fr}
\affiliation{Sorbonne Universit\'e, CNRS, PHENIX, F-75005 Paris, France}

\author{Maria Barbi}
\email{maria.barbi@sorbonne-universite.fr}
\affiliation{Sorbonne Universit\'e, CNRS, LPTMC, F-75005 Paris, France}

\date{\today}

\begin{abstract}

We aim at calculating an explicit expression for the finite-size probability distribution of the radius of gyration $R$ of an Interacting Self-Avoiding Walk (ISAW), as a function of chain length $N$ and monomer-monomer interaction energy $\varepsilon$.
We first derive the explicit free energy expression for a non-interacting Self-Avoiding Walk, introducing a new natural scale variable $t = \rho^g$ expressed as a power of the density $\rho$. Then, thanks to a cumulant expansion approach introduced by Lhuillier, Victor and coworkers, we extend it to the interacting case, capturing both the coil-globule transition and finite-size corrections to scaling, including entropic and surface-energy contributions. The radius of gyration distribution determined by the new free energy expression is then compared, using Bayesian inference to estimate the model parameters, to results from extensive Monte Carlo simulations of three-dimensional ISAWs, showing excellent agreement except very close to the $\Theta$-point, where finite chain lengths limit the accessible scaling regime. This validates our resulting free energy expression and allows us, as an application, to construct a phase diagram of the coil-globule transition. 
\end{abstract}

\maketitle

\section{Introduction}

Polymers - materials made of long molecules or macromolecules - are ubiquitous in soft matter: They are used in paints and in food products as shear thinning and gelifying agents; they can be found in soils, as a product of the biodegradation of organic matter~\cite{Tochiyama2004};
they are used in nanotechnologies (nanomaterials coated with polyelectrolyte brushes~\cite{Dobrynin2008}, electrochemical devices, solar cells~\cite{Chong2011});
last but not least, in living organisms genetic information is carried by very long DNA macromolecules, which are associated with DNA-binding proteins in a complex called chromatin that can again be assimilated to a polymeric material.
An important part of polymer studies consists in identifying the essential factors that control the structural, thermodynamic and dynamic properties of their constituent macromolecules, either in solution, in a melt or in complex environments, and to assess their relevance by comparing experimental results with model predictions.
Polymer theories have been successful in using the scale invariance of macromolecules to determine scaling relationships between the three-dimensional size, typically quantified by the \emph{radius of gyration} of the chains (see \autoref{eq:RG1}) and its dependence on their degree of polymerization (or \textit{number of monomers}, $N$).
Typically, the observed scaling allows discriminating between swollen or \textit{coil} polymer conformations and the collapsed, dense conformations called \textit{globules} that can be observed depending on the physico-chemical properties of polymer and the surrounding solvent~\cite{Grosberg1994}.
The radius of gyration can be related to various experimental measurements, from neutron diffraction spectra to self-diffusion coefficients.
Interestingly, recently developed super-resolution imaging techniques provide series of images of single macromolecule conformations, from which the whole \emph{distribution} of the radius of gyration can be extracted.

Different approaches are possible in order to describe polymer physics, from atomistic to highly coarse-grained models, depending on the description aim and system features~\cite{Spakowitz2019}.
Our modeling framework is that of self-attracting linear homopolymers, hereafter named ISAW for \emph{Interacting Self-Avoiding Walk}s.
In the limit of infinitely long chains, ISAW polymers undergo a true phase transition, called \emph{coil-globule transition}, as the monomer-monomer attractive interaction becomes strong enough relative to the thermal energy~\cite{DeGennes1975,Maffi2012}.
In the case of real polymers, the limited number $N$ of monomers leads however to \emph{finite-size effects}, which induce strong deviations from the infinite scaling behavior \cite{Grassberger1995,Rissanou2006,Allahverdi2010}.
Despite a number of studies~\cite{Grassberger1995, Imbert1996, Imbert1997}, there is no theoretical expression for the free energy of a finite-size ISAW that is actually reliable in the whole range  of states from coil to globule. The derivation is indeed particularly challenging, since the same free energy formula should be able to predict the fluctuations of the radius of gyration for conditions where these fluctuations are qualitatively very different (\eg{} in the coil or in the globule) and may be driven by various thermodynamic forces (\eg{} conformational entropy and enthalpic terms).

The main purpose of this work is to show how to obtain an explicit and accurate formula for this free energy as a function of its instant radius of gyration $R$, of the effective monomer-monomer interaction energy and of the polymer length $N$. The order parameter can also be replaced by equivalent quantities as for example the monomer density $\rho = N / R^d$ or more generally any power of the density, $\rho^\eta$, with arbitrary $\eta$.
To obtain the new free energy expression, we proceed along the same lines as in the series of papers by Lhuillier, Victor and colleagues~\cite{Lhuillier1988,Victor1990, Victor1994, Imbert1996, Imbert1997}, starting from the free energy of a (non interacting) \emph{ Self-Avoiding Walk} (SAW), with the introduction of a cumulant expansion.

The paper is organized as follows. We first introduce the ISAW model and the coil-globule transition in \autoref{sec:ISAW}, where we recall the expected scaling laws for the radius of gyration. In \autoref{sec:P_SAW}, we first address the problem of determining the finite-size distribution of the radius of gyration for a SAW, i.e. in the limit where no attractive forces are present, and in the infinite-size limit. Following scaling arguments, we postulate a free energy expression, further used to built the more general form in section \autoref{sec:P_ISAW}.  We then revisit the coil–globule transition within a Landau-like free-energy framework, identifying the $\Theta$-point as the vanishing of the effective two-body interaction and discussing the associated tricritical scaling laws (\autoref{subsec:coil-globule}).
To allow for generalization and comparison to finite-size system, we then discuss in  \autoref{sec:Finite-size corrections to scaling} the  finite-size corrections to scaling, including entropy and surface energy terms.
We finally validate in our analytical expression \autoref{sec:numerical}  by comparing with Monte Carlo simulation results, showing a very accurate agreement except very close to the critical value, due to a limited polymer size $N$. As an application, we also present a phase diagram of the coil-globe transition based on the theoretical expression, before drawing our {Conclusions}.

\section{From SAW to ISAW}
\label{sec:ISAW}

In order to  model the conformation of a self-interacting real chain, it is necessary to take into account, first of all, the excluded volume interactions, preventing the chain from overlapping with itself.
To model self-avoidance, one possibility is in principle to start from a random walk, whose free energy is known, and to add a repulsive interaction to avoid overlapping and links crossing~\cite{Orland1994}.
However, if excluded volume can be introduced by a hard sphere potential, there is no analytic solution to prevent link crossings at the moment.
A second choice, first introduced by Lhuillier~\cite{Lhuillier1988} then adopted by Victor~\cite{Victor1990, Victor1994, Imbert1996, Imbert1997}, is to start directly from a self-avoiding walk (SAW), whose free energy is not known exactly, but whose behavior can be studied through simulations.
In this way, all the features of self-avoidance are inherently ensured.
In this section, we revisit the latter approach by a new, more straight-forward derivation based on the system free energy.
On this basis, we will then be able to extend it to the case of ISAW by introducing an energy gain for monomer-monomer contacts.

We first model a polymer as a self-avoiding walk made of $N$ identical monomers, located at positions $\boldsymbol r_{i}$ (for $i=1,\dots,N$) on a lattice (see \autoref{sec:simulations}).
A standard way to quantify the spatial extent of a single polymer chain in a given configuration is the instant radius of gyration $R$, defined as the standard deviation of the instant distribution of the monomer positions, \ie{} :
\begin{equation} \label{eq:RG1}
	R^2
	    = \frac{1}{N}\sum_{i=1}^{N} \left( \boldsymbol r_{i} - \boldsymbol G \right)^2
	    = \frac{1}{N}\sum_{i=1}^{N} \boldsymbol r_i^2 - \boldsymbol G^2,
\end{equation}
with 
\begin{equation}
\boldsymbol G = \frac{1}{N}\sum_{i=1}^{N} \boldsymbol r_i \,.    
\end{equation}
It is common to statistically characterize the average behavior of a polymer of $N$ monomers by means of the \emph{mean} radius of gyration,
\begin{equation} \label{eq:meanRG}
	\bar R = \sqrt{\langle R^2 \rangle} \,,
\end{equation}
where the average $\langle \cdot \rangle$ is performed over the ensemble of conformations of the given polymer.

When attraction is included, the polymer chain is modelled by an ISAW (Interacting Self-Avoiding Walk), i.e. a SAW (self-avoiding chain, here on a cubic lattice) with attractive interactions between the nearest neighbour monomers on the lattice (but not consecutive along the chain). The energy of the chain is $E=-mJ$, with $m$ the number of contacts (number of nearest neighbour pairs on the lattice) and $-J$ the interaction energy per contact. At a given temperature $T$, the only relevant parameters are the size $N$ of the polymer and the dimensionless ratio $\varepsilon = J/k_BT$, accounting for the relative strength of the effective monomer-monomer interactions with respect to temperature.

Depending on $\varepsilon$, two different folding modes have been predicted and measured~\cite{Nishio1979} for a single polymer chain at equilibrium and in the large-$N$ limit.
In good solvent (low $\varepsilon$), the favorable interaction with the solvent leads to an effective repulsion between monomers.
Hence, the polymer expands into a decondensed, disordered state called \emph{coil}, described as a \emph{self-avoiding walk} (SAW).
In poor solvent (high $\varepsilon$), monomer-monomer attractions become predominant, and the polymer collapses into a state called \emph{globule}.

For infinite-size polymers a continuous phase transition between these two phases is observed at a specific temperature called $\Theta$ (\emph{theta}) temperature or $\Theta$-\emph{point}, or equivalently at $\varepsilon_\Theta = J/k_B\Theta$~\cite{DeGennes1975,Maffi2012}.
At this point, the polymer follows the same scaling laws as a random walk (RW) while preserving self-avoidance (also referred to as $\Theta$-polymer).
For finite-size polymers, the transition is replaced by a continuous crossover~\cite{Grassberger1995,Rissanou2006,Allahverdi2010} (however, in the following, the term \emph{transition} will also be used for finite-size polymers, for convenience).
The energy of appearance of this crossover becomes a function $\varepsilon_\Theta(N)$ of the polymer length $N$ and it is always higher than $\lim_{N \to \infty}\varepsilon_\Theta(N) = \varepsilon_\Theta$, until coinciding in the infinite-size limit~\cite{Grassberger1995,Vogel2007}.
This effective $N$-dependent transition energy is a consequence of the connectivity of the system, that induces strong cooperation effects in the globule: the smaller the number of monomers, the larger the energy per contact required to stabilize the globular conformations.
Interestingly, in finite-size the phase transition critical line (defined \eg{} by the vanishing of the second virial coefficient, or by a divergent specific heat) dissociates from the condition of RW scaling, that generally appears at lower energy~\cite{desCloizeaux1991, Foldes2021}.

Due to their different folding features, the two different regimes induce two different scaling laws for the dependence of the volume occupied by the polymer with the number of monomers $N$.
In particular, the mean radius of gyration follows in each regime a power law as a function of $N$,
\begin{equation}
\bar R \propto N^\psi
\end{equation}
where the scaling exponent $\psi$ depends on the folding regime and takes one of the values summarized in \autoref{tab:exponents}.
\begin{table}[ht]
\bgroup
\def\arraystretch{1.4}
\begin{tabular}{|l|cc|}
\hline
  & Coil (SAW)$\;\;$ 
  & Globule \\
\hline
Condition: & $\varepsilon < \varepsilon_\Theta(N)$ 
& $ \varepsilon_\Theta(N) <\varepsilon$ \\
Scaling exponent: & $\nu$ 
& 1/3 \\ 
\hline
\end{tabular}%
\egroup
\caption{
    \label{tab:exponents}
    Summary of $\varepsilon = J/k_BT$ conditions and scaling exponents expected for the two typical polymer folding states in the thermodynamic limit.
    The exponent $\nu \approx 0.588$ is the Flory's exponent. The (infinite-size limit) critical value is $\varepsilon_\Theta \simeq 0.27$.
}
\end{table}

Finally, it is useful for our further development to mention that a polymer can adopt a stretched arrangement, typically under the action of an external force.
In this limit, the scaling law $\bar R \propto N^\psi$ becomes compatible with a scaling exponent $\psi =1$.
Depending on the interactions, a macromolecule can thus have from very condensed to very stretched conformations.
Three conformation categories can be introduced, even in finite-size, with reference to the large $N$ limit behavior: globules, swollen coils, and stretched.

\section{Distribution of the radius of gyration for a SAW}
\label{sec:P_SAW}

We aim at determining the finite-size distribution of the radius of gyration for a SAW, that we note ${\cal P} (R \big\vert 0, N)$, where "$0$" stays for $\varepsilon = 0$ indicating that no attractive energy is taken into account.
From a geometrical point of view, a SAW can take any (self-avoiding) conformation, from very condensed to very stretched ones.
Depending on physical conditions, the weight of globular, swollen, or stretched conformations in the corresponding distribution density can vary considerably, making alternately some of them very unlikely, others overwhelmingly represented.
However, when focusing on the finite-size \emph{distribution} of the radius of gyration, ${\cal P} (R \big\vert 0, N)$, it is possible to look for a global expression allowing for retrieving the good properties for the limiting cases.
\citeauthor{Imbert1997} have shown indeed that ${\cal P} (R \big\vert 0, N)$ has a different scaling behavior in each of these three classes~\cite{Imbert1997}.
As these different scalings are actually three limiting cases of a unique underlying distribution, there must be some strong matching conditions satisfied, as first shown by Lhuillier~\cite{Lhuillier1988}.

The stretched conformations are characterized by translational invariance 
and can be seen as a succession of independent parts due to the absence of excluded volume constraints between the segments \cite{Imbert1997}.
Consequently, the free energy of such a class of conformations is extensive and depends on its linear density $\lambda = N / R$ through a function $\mathcal{S}$.
We \emph{assume} that $\mathcal{S}$ is a power law of prefactor $A_S$ and power $s$:
\begin{equation} 
\beta F_S (R \big\vert 0, N) \sim N \mathcal{S}(\lambda) = N A_S \cdot \lambda^s \,.
\end{equation} 
Globular conformations are homogeneous, which means that the local monomer density does not depend on the position within the occupied region and coincides with the overall density $\rho = N / R^d$, where $d$ is the dimension of the space.
This implies that the free energy of such a conformation class 
is extensive and depends on its volume density $\rho = N / R^d$ through a function $\mathcal{G}$.
Within the low density limit ($\rho \to 0$), we \emph{assume} that this function $\mathcal{G}$ is a power law of prefactor $A_G$ and power $g$:
\begin{equation} 
\beta F_G (R \big\vert 0, N)  \sim N \mathcal{G}(\rho) = N A_G \cdot \rho^g \,.
\end{equation} 

Finally, the swollen coil conformations are scale invariant, with fractal dimension $1/\nu$, where $\nu \approx 0.588$ is again the Flory exponent.
Hence, the free energy of such a conformation class only depends on the scale variable $\phi = N/R^{1/\nu}$ through a function $\mathcal{C}$.
In order to guarantee that one can move continuously from a stretched conformation to a low-density globular conformation through a coil, we \emph{assume} here that $\mathcal{C}$ is a sum of power laws (of prefactors $A_S$, $A_G$ and exponents $c_s$, $c_g$), so that we can write
\begin{equation} 
\beta F_C (R \big\vert 0, N) \sim \mathcal{C}(\phi) 
 = A_S \cdot \phi^{c_s} + A_G \cdot \phi^{c_g} \,.
\end{equation} 
At each of the two limits, alternately, one of the two terms will prevail over the other, allowing connection with the corresponding configuration classes while ensuring both continuity and the correct scaling law.
The continuity condition is indeed imposed by the fact that the three classes are nothing but a partition of a same self-avoiding walk conformation ensemble.

At the stretch-coil limit, the expression of the two free energies must coincide: 
\begin{equation} 
A_S \cdot N \left( \frac{N}{R} \right)^s = A_S \cdot \left( \frac{N}{{R}^{1/\nu}} \right)^{c_s} 
\end{equation} 
To do this, it is necessary to match the exponents of $N$ and $R$, on both sides, leading to the system: 
\begin{equation} 
\begin{cases} 
1 + s & = c_s \\\ 
\nu s & = c_s 
\end{cases} 
\implies 
\begin{cases} 
c_s & = -\nu / (1 - \nu) \\\ 
s & = -1 / (1 - \nu) 
\end{cases} 
\end{equation} 

By introducing the Fisher-Pincus exponent~\cite{DeGennes1979} 
$\delta = {1}/{(1 - \nu)}$, we can rewrite the exponents as $c_s = -\nu\delta$ and $s = -\delta$.

Similarly, at the limit between a low-density globule ($\rho \to $0) and a coil, the expression of the two free energies must coincide: 
\begin{equation} 
    A_G \cdot N \left( \frac{N}{R^d} \right)^g = A_G \cdot \left( \frac{N}{R^{1/\nu}} \right)^{c_g} 
\end{equation} 
Equating the $N$ and $R$ exponents gives rise to the system: 
\begin{equation}
\label{eq:exposants-coil-globule}
    \begin{cases}
        1 + g & = c_g \\
        \nu d g & = c_g
    \end{cases}
    \implies
    \begin{cases}
        c_g & = \nu d / (\nu d - 1) \\
        g & = 1 / (\nu d - 1)
    \end{cases}
\end{equation}

In the globule phase, a natural scale variable, $t = \rho^g$, appears: it can be interpreted as a renormalized density, taking into account self-avoidance, with $g = 1 / (\nu d - 1)$. Note that the {coil-globule} transition occurs in the low density limit $\rho \to 0$, hence for $t \to 0$. In this limit the free energy $\beta {\cal F} (R \big\vert 0, N)$, or equivalently $\beta F (t \big\vert 0, N)$ when given as a function of the renormalized density t, has therefore the following asymptotic expansion 
\begin{equation}
\label{eq:distribution-P_SAW-de-t}
\beta F (t \big\vert 0, N) = N (A_G \cdot t + o(t))
\end{equation}

We assume that the free energy $\beta F (t \big\vert 0, N)$ is an analytic function of $t$ in the vicinity of zero, so that
\begin{equation} 
\beta F (t \big\vert 0, N) = N (A_G \cdot t + A^{'}_G \cdot t^2 + o(t^2))
\end{equation}

To justify the assumption that the free energy $\beta F(t \mid 0, N)$ is analytic in $t = \rho^g$ (with $g = 1/(\nu d - 1)$) in the low-density limit $t \to 0$, we note that in this regime, excluded-volume interactions become perturbative, and the system effectively behaves as an ideal chain with analytic corrections. This is consistent with the virial expansion framework for dilute polymer solutions, where the free energy admits a power-series expansion in density (or renormalized density $t$) with integer exponents, as discussed in de Gennes’ scaling theory of polymers~\cite{DeGennes1979}. Furthermore, since $t = 0$ does not correspond to a critical point for $\varepsilon = 0$ (the coil-globule transition requires attractive interactions), the Lee-Yang theorem~\cite{LeeYang1952} ensures the absence of singularities in this limit. The choice of $t$ as a renormalized variable, dictated by the matching conditions between the stretched, coil, and globule regimes (as established by Lhuillier~\cite{Lhuillier1988} and \citeauthor{Imbert1997}~\cite{Imbert1997}), ensures that critical scaling effects are absorbed, leaving a smooth, analytic free-energy landscape. This ansatz is further supported by our numerical simulations, which exhibit excellent agreement with the leading-order terms of the expansion (see \autoref{sec:numerical}).

This free energy is purely entropic, which is usual with self-avoiding walks (each conformation has zero energy), hence
\begin{equation}
\beta {\cal F} (R \big\vert 0, N) = -{\cal S} (R \big\vert 0, N)/ k_B
\end{equation}
Finally, we obtain the distribution of $R$ as
\begin{equation}
\label{eq:PofRg2}
    {\cal P} (R \big\vert 0, N) 
= \frac{1}{{\cal Z} (0, N)}
\exp{\big(-\beta {\cal F} (R \big\vert 0, N) \big)} \,,
\end{equation}
where the partition function ${\cal Z} (0, N)$ is equal to the total number of conformations of a SAW of size $N$, usually denoted by $\aleph_N$. We know that this number is given by the formula
\begin{equation}
\label{eq:aleph_N}
    \aleph_N = \Lambda \mu^N N^{\gamma-1}
\end{equation}
where $\Lambda$ is a (lattice-dependent) constant, $\mu$ the effective connectivity of the lattice and $\gamma$ the enhancement exponent~\cite{DeGennes1979}.

Finite-size corrections to scaling will be discussed in  \autoref{sec:Finite-size corrections to scaling} These corrections are significant for the numerical evaluation of  ${\cal P} (R \big\vert \varepsilon, N)$. However they become negligible as $N \to \infty$ and will not be considered in the theoretical analysis of the coil-globule transition in the thermodynamic limit that we perform in \autoref{subsec:coil-globule}.

\section{Distribution of the radius of gyration for an ISAW}
\label{sec:P_ISAW}

\subsection{Formal cumulant expansion}

When considering an ISAW, the energy of the chain in a given conformation becomes \mbox{$E=-mJ$}, with $m$ the number of contacts (number of nearest neighbour pairs on the lattice). Hence, this energy enters in the expression of the distribution of the radius of gyration though a Boltzmann factor. Explicitly,
the distribution of the radius of gyration $R$ at any $\varepsilon$ can be written:
\begin{equation}
\label{eq:PdeR}
    {\cal P} (R \big\vert \varepsilon, N) = \frac{{\cal Z} (R \big\vert \varepsilon, N)}{{\cal Z}(\varepsilon, N)}
\end{equation}
where
\begin{equation}
\label{eq:ZdeR}
    {\cal Z} (R \big\vert \varepsilon, N) = \sum_{\mathscr{C}(R)} e^{\varepsilon m_{\mathscr{C}(R)}}
\end{equation}
is the partition function of a chain of size $N$ and radius $R$ at $\varepsilon$, and
\begin{equation}
\label{eq:Z}
    {\cal Z} (\varepsilon, N) 
= \sum_\mathscr{C} e^{\varepsilon m_\mathscr{C}}
\end{equation}
is the partition function of the chain of size $N$ at $\varepsilon$. The notation $m_\mathscr{C}$ denotes the number of contacts in the $\mathscr{C}$ conformation and $m_{\mathscr{C}(R)}$ the number of contacts in the ${\mathscr{C}(R)}$ conformation. As before,  ${\cal Z} (0, N) = \aleph_N = = \Lambda \mu^N N^{\gamma-1}$ is the total number of conformations of a SAW of size $N$, usually denoted by $\aleph_N$.

Moreover, we have the following relations between the partition functions and the corresponding free energies:
\begin{equation}
\label{eq:ZdeR-FdeR}
    \ln {\cal Z} (R \big\vert \varepsilon, N)
= -\beta F (R \big\vert \varepsilon, N)
\end{equation}
and
\begin{equation}
\label{eq:Z-F}
    \ln {\cal Z} (\varepsilon, N)
= -\beta F (\varepsilon, N).
\end{equation}

The partition functions ${\cal Z} (R \big\vert \varepsilon, N)$ and ${\cal Z} (\varepsilon, N)$ can in turn be written as
\begin{equation}
    {\cal Z} (R \big\vert \varepsilon, N)
= {\cal Z} (R \big\vert 0, N) \ \langle{e^{m \varepsilon}}\rangle_{0,N,R} 
\end{equation}
and
\begin{equation}
    {\cal Z} (\varepsilon, N)
= {\cal Z} (0, N) \ \langle{e^{m \varepsilon}}\rangle_{0,N} 
\end{equation}
where$\langle{\cdot}\rangle_{0,N}$  denotes the mean in the canonical set for an SAW of size $N$ (i.e. an ISAW of size $N$ at $\varepsilon=0$) and $\langle{\cdot}\rangle_{0,N,R}$  the same mean \textit{restricted} to conformations of radius $R$. We deduce the following relations for the free energies:
\begin{equation}
  \beta F (R \big\vert \varepsilon, N)
= \beta F (R \big\vert 0, N) - \ln{\langle{e^{m \varepsilon}}\rangle_{0,N,R}}
\end{equation}
and
\begin{equation}
 \beta F ( \varepsilon, N)
= \beta F ( 0, N) - \ln{\langle{e^{m \varepsilon}}\rangle_{0,N}}
\end{equation}

One recognizes, in the second members of these equations, the generating function $K_\varepsilon(R)$ (resp. $K_\varepsilon$) of the cumulants $\kappa_n(R)$ of the distribution of the number of contacts at a fixed $R$ (resp. of the cumulants $\kappa_n$ of the distribution of the number of contacts):
\begin{equation} 
\label{eq:cumulant-of-R-generating-function} 
K_\varepsilon(R) = \ln \langle{e^{\varepsilon m}}\rangle_{0,N,R}
= \sum_{n=1}^{+\infty} \kappa_n(R) \frac{\varepsilon^n}{n!}
\end{equation} 
and
\begin{equation} 
\label{eq:fonction-generatrice-des-cumulants} 
K_\varepsilon = \ln \langle{e^{\varepsilon m}}\rangle_0
= \sum_{n=1}^{+\infty} \kappa_n \frac{\varepsilon^n}{n!}.
\end{equation} 
Eventually, we have:
\begin{equation} 
\label{eq:free-energy-expansion-of-R} 
 \beta F (R \big\vert \varepsilon, N)
= \beta F (R \big\vert 0, N) - \sum_{n=1}^{+\infty} \kappa_n(R) \frac{\varepsilon^n}{n!}
\end{equation} 
and
\begin{equation} 
\label{eq:free-energy-expansion} 
 \beta F ( \varepsilon, N)
= \beta F ( 0, N) - \sum_{n=1}^{+\infty} \kappa_n \frac{\varepsilon^n}{n!}
\end{equation}
The cumulants $\kappa_n(R)$ and $\kappa_n$ are independent of $\varepsilon$. Remarkably, then, the free energy of an ISAW is given in terms of the statistical properties of a SAW.

Let us return to the distribution of the radius of gyration $R$ at any temperature $T$ given in the equation \autoref{eq:PdeR}. Using the equations \autoref{eq:ZdeR-FdeR} and \autoref{eq:Z-F} we obtain:
\begin{equation}
    {\cal P} (R \big\vert \varepsilon, N) 
= \exp{\big(-(\beta F (R \big\vert \varepsilon, N) - \beta F ( \varepsilon, N))\big)}
\end{equation}
then by using the equations \autoref{eq:free-energy-expansion-of-R} and \autoref{eq:free-energy-expansion} we finally find:
\begin{equation}
\label{eq:distribution-P_ISAW-de-R}
    {\cal P} (R \big\vert \varepsilon, N) 
= {\cal P} (R \big\vert 0, N) \exp{\big(\sum_{n=1}^{+\infty} (\kappa_n(R)-\kappa_n) \frac{\varepsilon^n}{n!}\big)}
\end{equation}
where
\begin{equation}
\label{eq:distribution-P_SAW-de-R}
{\cal P} (R \big\vert 0, N) = \exp{\big(-(\beta F (R \big\vert 0, N) - \beta F ( 0, N))\big)}.
\end{equation}
\autoref{eq:distribution-P_ISAW-de-R} gives a formal expression for the ISAW distribution of the radius of gyration. In order to obtain explicit dependencies in $R$ or, equivalently, any power of the density $\rho$, we now need to investigate  the cumulants scaling properties.

\subsection{Cumulants scaling laws}

The cumulants $\kappa_n(R)$ of the distribution of the number of contacts at a fixed $R$ have similar scaling behavior to that of entropy. Indeed, let us first consider the first cumulant, $\kappa_1(R)$, which is equal to the average number of contacts of a SAW:
\begin{equation}
    \kappa_1(R)
    = \langle{m}\rangle_0 (R) = a_1N + f(\phi)
\end{equation}
where $\langle{\cdot}\rangle_0$ refers to the average in the canonical ensemble at $\varepsilon = 0$ and $\phi = N/R^{1/\nu}$.
The first term $a_1N$ corresponds to the contacts between immediate neighbors along the polymer chain; the second, $f(\phi)$, is the average number of contacts between distant monomers along the chain. In the coil phase, the latter only depends on the scale variable $\phi$,  whereas in the globule phase, it depends on the density $\rho = N / R^d$ and is proportional to $N$. Consequently, the term $f(\phi)$ writes $Ng(\rho)$. Finally, for globular conformations of density $\rho \to 0$ the two expressions must coincide, so that we can write
\begin{equation}
    f(\phi) = Ng(\rho)
\end{equation}
The functions $f$ and $g$ are power laws and we therefore find the same exponents $g$ and $c_g$ as in the equation \autoref{eq:exposants-coil-globule}. We finally obtain an asymptotic expansion in the renormalized density $t$ similar to that already obtained for the entropy of a SAW:
\begin{equation}
\kappa_1(R)
    = \langle{m}\rangle_0 (R) = N (a_1 + B_G \cdot t + o(t))
\end{equation}

The higher order cumulants have a very similar asymptotic development. To show this, we begin by taking the decomposition of the number of $m$ contacts into two categories: contacts between neighbouring monomers along the chain (contacts \emph{in cis} whose number is $m_1$) or between distant monomers along the chain (contacts \emph{in trans} whose number is $m_2$):
\begin{equation}
    m = m_1 + m_2.
\end{equation}
In the coil phase, the distribution of the number $m_1$ of \emph{in cis} contacts depends weakly on the radius of gyration, whereas the distribution of the number $m_2$ of \emph{in trans} contacts is a function of $\phi = N/R^{1/\nu}$. 
 In the globule phase, the distribution of the number $m_1$ of \emph{in cis} contacts again depends weakly on the radius of gyration, whereas the distribution of the number $m_2$ of trans contacts depends on the density $\rho = N / R^d$ and is proportional to $N$.
All cumulants, whatever their order, are extensive \cite{RodriguezTsallis2010}. We therefore propose the following asymptotic expansion:
\begin{equation}
\kappa_n(t)
    = N (a_n + B_n \cdot t + B^{'}_n \cdot t^2 + o(t^2)).
\end{equation}

By inserting in \autoref{eq:free-energy-expansion-of-R} the complete expression of the free energy, we get
\begin{equation} 
 \beta F (t \big\vert \varepsilon, N)
= \beta F (t \big\vert 0, N) - \sum_{n=1}^{+\infty} \kappa_n(t) \frac{\varepsilon^n}{n!},
\end{equation} 
and, consequently,
\begin{align}
 \beta F (t \big\vert \varepsilon, N)
&= N (A_G \cdot t + A^{'}_G \cdot t^2 + o(t^2)) \nonumber
\\&- \sum_{n=1}^{+\infty} N (a_n + B_n \cdot t + B^{'}_n \cdot t^2 + o(t^2)) \frac{\varepsilon^n}{n!}.
\end{align}

Reordering this summation over $t$ we get
\begin{align} 
 \beta F (t \big\vert \varepsilon, N)
&= N \biggl\{ - \sum_{n=1}^{+\infty} a_n \frac{\varepsilon^n}{n!} + \biggl[ A_G - \sum_{n=1}^{+\infty} B_n \frac{\varepsilon^n}{n!} \biggr] \cdot t \nonumber\\
&+ \biggl[ A^{'}_G - \sum_{n=1}^{+\infty} B^{'}_n \frac{\varepsilon^n}{n!} \biggr] \cdot t^2 + o(t^2))  \biggr\} + o(N)
\end{align} 

Hence, it is possible to rewrite the free energy as
\begin{equation} \label{eq:free-energy-expansion-of-t-explicit}
   \beta F (t \big\vert \varepsilon, N)
= N \bigl\{ a_0(\varepsilon) + a_1(\varepsilon) \cdot t + a_2(\varepsilon) \cdot t^2 + o(t^2)  \bigr\} + o(N) 
\end{equation}
where the $\varepsilon$-dependent parameters $a_0$, $a_1$, $a_2$ are independent of $N$ and $t$.

\subsection{Identifying the $\Theta$-point in the free-energy expansion
}
\label{subsec:coil-globule}

The free-energy expansion \autoref{eq:free-energy-expansion-of-t-explicit} is reminiscent of the low-density virial expansion, with the renormalized density $t=\rho^{1/(\nu d-1)}$.
In this analogy, the coefficients $a_1$ and $a_2$ play the role of the second and third virial coefficients, respectively. In particular, the sign of $a_1$ determines whether repulsive ($a_1>0$) or attractive ($a_1<0$) interactions dominate. We therefore expect $a_1(\varepsilon)$ to vanish at a critical value $\varepsilon=\varepsilon_\Theta$, defining the $\Theta$ temperature (or $\Theta$-\emph{point}).

As a simple approximation, $a_1(\varepsilon)$ is taken
to vanish linearly,
$a_1(\varepsilon)\propto\varepsilon-\varepsilon_{\Theta}$.
This approximation is in good practical agreement with the numerical simulations presented in \autoref{sec:numerical} over the range of $\varepsilon$ explored.

We therefore introduce the reduced distance to the $\Theta$-point,
\begin{equation}
    \tau=\frac{\varepsilon-\varepsilon_\Theta}{\varepsilon_\Theta},
\end{equation}
so that $a_1(\varepsilon) = a_1' \tau \propto\tau$ in the vicinity of the $\Theta$-point.

This control parameter is equivalent, up to a sign, to the conventional reduced temperature ${(T-\Theta)}/{\Theta}$. The next coefficient, $a_2(\varepsilon)$, which plays the role of an effective three-body interaction, is expected to remain finite and positive at $\varepsilon=\varepsilon_\Theta$. Its positivity ensures the stability of the free-energy expansion when the quadratic term in $t$ becomes dominant at the $\Theta$-point, so to prevent collapse of the polymer at the transition and to stabilize the globule phase. The asymptotic expansion \autoref{eq:free-energy-expansion-of-t-explicit} of the free energy $\beta F (t \big\vert \varepsilon, N)$ therefore becomes in the vicinity of $\varepsilon=\varepsilon_{\Theta}$
\begin{equation} \label{eq:free-energy-expansion-of-t-around-theta}
     \beta F (t \big\vert \varepsilon, N)
= N \bigl\{ a_0(\varepsilon_{\Theta}) + a_1'\tau \cdot t + a_2(\varepsilon_{\Theta}) \cdot t^2 + o(t^2)  \bigr\} + o(N) 
\end{equation}
We can build two scaling variables: $\hat{t}=N^{\frac{1}{2}}t$ and $\hat{\tau}=N^{\frac{1}{2}}\tau$ which enable to rewrite the singular part of the free energy so to get
\begin{align}
\label{eq:with_hat_t}
     \beta F (t \big\vert \varepsilon, N)
     = \beta \hat{F} (\hat{t},\hat{\tau})
&= N a_0(\varepsilon_{\Theta})  
+ a_1'\hat{\tau} \cdot \hat{t} + a_2(\varepsilon_{\Theta}) \cdot \hat{t}^2\nonumber \\
& + o(\hat{t}^2) + o(N).
\end{align}
The free energy depends on the two scaling combinations $\hat t$ and $\hat\tau$, revealing the tricritical nature of the $\Theta$-point (as is well known since de Gennes~\cite{DeGennes1979}) with crossover exponent $\Phi={1}/{2}$.

\section{Finite-size corrections to scaling}
\label{sec:Finite-size corrections to scaling}

{
The functions $a_i(\varepsilon)$ in \autoref{eq:free-energy-expansion-of-t-explicit} can be fitted by comparing the predicted distribution of gyration radii \autoref{eq:PofRg2} to numerical data obtained by Monte Carlo simulations.
To this aim, we ran extensive on-lattice simulations at different values $N$ and $\varepsilon$, as described in \autoref{sec:numerical}. However, we first need to specify the finite-size corrections to scaling. This is the subject of the present section.
}

\subsection{Entropy term}
\label{Entropy terms}

${\cal P} (R \big\vert 0, N)$ is the distribution of the gyration radius $R$ of an SAW of size $N$. We will now denote it more explicitly as ${\cal P}^{SAW}_{N} (R)$ or, equivalently, $P^{SAW}_{N} (t)$ when given as a function of the renormalized density $t$.
We show here that it is necessary to introduce additional finite-size effects to numerically fit this probability distribution function.

Let $\aleph_N(t)$ denote the number of conformations with a given renormalized density $t$ that a SAW of size $N$ can adopt. This number can be given a similar expression as  \autoref{eq:aleph_N}~\cite{Owczarek1993},
\begin{equation}
\aleph_N(t) = \Lambda(t) \mu(t)^N N^{\gamma_g - 1},
\end{equation}
where $\mu(t)$ is the connectivity of a globule of density $t$, $\gamma_g$ is the enhancement exponent for globular conformations and $\Lambda(t)$ is a factor that depends on both the lattice type and the density.

At high density, i.e. $t =O(1)$, the probability of observing a (globular) conformation of density $t$ reads 
\begin{equation}
{p}^{SAW}_{N} (t) = \frac{\aleph_N(t)}{\aleph_N} = \frac{\Lambda(t)}{\Lambda} \left( \frac{\mu(t)}{\mu} \right)^N N^{\gamma_g - \gamma} \,.
\end{equation} 
This probability must depend on $N$ as $p_N \propto \exp(-N \mathcal{G}(t))$, which allows for the identification 
\begin{equation} 
\frac{\mu(t)}{\mu} = e^{-\mathcal{G}(t)}. 
\end{equation} 
Following \autoref{eq:distribution-P_SAW-de-t} we get
\begin{equation} 
\mathcal{G}(t) = A_G \cdot t + o(t). 
\end{equation}

\citeauthor{Imbert1997}~\cite{Imbert1997} have shown that
\begin{equation}
\mathcal{G}(t) - A_G \cdot t \propto t^2
\end{equation}
so that the asymptotic expansion can be extended to
\begin{equation} 
\mathcal{G}(t) = A_G \cdot t + A^{'}_G \cdot t^2 + o(t^2), 
\end{equation}
again in agreement with the fact that $\mathcal{G}(t)$ is an analytic function of $t$ in the vicinity of zero.

Moreover, these authors have speculated that the factor 
${\Lambda(t)}/{\Lambda} $ also follows a power law with an exponent $-c$ : ${\Lambda(t)}/{\Lambda} \propto t^{-c}$.
They showed that this exponent has to satisfy the relationship $c = 1 + \gamma - \gamma_g$, and measured the value $c \approx $1.13 using simulations.

The free energy of a SAW can therefore be written as a function of the density $t$ as 
\begin{equation} \label{eq:free-energy-saw} 
\beta F^{SAW}_{N} (t) = A_G \cdot N t + A^{'}_G \cdot N t^2 + A_S \cdot \left( N t \right)^{-q} + c \ln(N t). 
\end{equation} 

This free energy is purely entropic, as usual with self-avoiding walks (each conformation has zero energy), hence
\begin{equation}
\beta F^{SAW}_{N} (t) = -S^{SAW}_{N} (t)/ k_B.
\end{equation}\\


As we will show later, the model, as it is defined through the free energy with the surface term that we will introduce in next subsection, is very satisfactory.
The distribution of gyration radius for each value of the parameters, as well as the dependence of the mean and the median of the gyration radius on $N$ and on $\varepsilon$ agree with the simulation results.
Nevertheless, there are some indications that this model is less effective very close to the coil-globule transition, inducing deviations for large $R$ values that, however, do not affect the median gyration radius, as expected.

By a careful trial and error steps, we identify that the main weakness of our model may come from the last term of \autoref{eq:free-energy-saw}, $c\ln N t$.
After several trials, we obtained a significant improvement of the agreement between the theory and the reference ISAW Monte Carlo simulations (described in the methods section), using the following function:
\begin{equation} \label{eq:corrected-cN}
    c_N(\varepsilon) = c_0(\varepsilon) + \frac{c_1(\varepsilon) }{N} + c_2(\varepsilon) \ln N\,,
\end{equation}
so that \autoref{eq:free-energy-saw} should now be written in the corrected form 
\begin{equation} \label{eq:corrected-free-energy-saw} 
\beta F^{SAW}_{N} (t) = A_G \cdot N t + A^{'}_G \cdot N t^2 + A_S \cdot \left( N t \right)^{-q}  + c_N(\varepsilon)\ln N t\,. 
\end{equation} 

We observed that the $c_N(\varepsilon)$ term smooths the transition from the coil to the globule distribution (where nevertheless a slight peak persists), at the cost of three additional parameters.

\subsection{Surface energy term}

\begin{figure*}[ht!]
    \centering
    \adjustbox{width=.95\textwidth}{\input{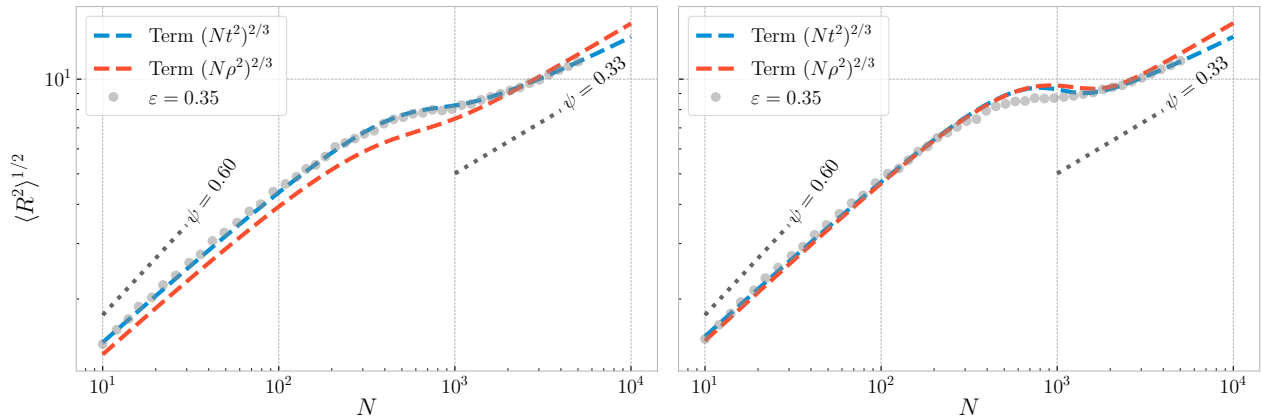}}
    \caption{
        Comparison of model predictions  with sampled distributions (dots) in medians (left) and in means (right), for the free energy $\beta F_\varepsilon(t) = a_1(\varepsilon) \cdot N t + a_2(\varepsilon) \cdot N t^2 + a_3(\varepsilon) \cdot (N t)^{-q} + c \ln N t $ supplemented by two different surface energy terms: ours $(N t^2)^{2/3}$ (\autoref{eq:corrected-free-energy-isaw-with-c}, blue dashed line) and Khokhlov's $(N \rho^2)^{2/3}$ (red dashed line).
    }
    \label{fig:cmp-a4}
\end{figure*}

The expansion on $t$ of the cumulant-generating function proposed above is still incomplete, due to the omission of the effect of surface tension, that is proved to be relevant for finite-size polymers~\cite{Rissanou2006}.
Surface tension appears in globular conformations and results in an energy penalty for monomers at the interface with the solvent, which is proportional to the surface area of the conformation.
In addition, \citeauthor{Rissanou2006} showed on simulation systems up to $N=10^4$ that the density is lower at the surface of the globular conformations than in the center of the globule, the low density region having a thickness of at least half the radius of the conformation~\cite{Rissanou2006}.
The effect is therefore far from negligible, leading to the necessity of introducing an appropriate surface energy term $\beta F_\sigma$.

A first attempt to model the surface energy has been proposed by \citeauthor{Lifshitz1978}~\cite{Lifshitz1978}, then supplemented by \citeauthor{Khokhlov1981}~\cite{Khokhlov1981} and included in \citeauthor{Grosberg1994}'s book~\cite{Grosberg1994} in the form $\beta F_\sigma \sim R^{d-1} {\tau_T}^2$, where ${\tau_T} = (T - \Theta) / \Theta$ is the relative deviation of the temperature $T$ from the temperature $\Theta$ of the coil-globule phase transition for an infinite-size polymer.
In \citeauthor{Khokhlov1981}'s theory, ${\tau_T}$ is proportional to $\rho$~\cite{Khokhlov1981}, so that his surface term can be written as $\beta F_\sigma \sim N^{(d-1)/d} \rho^{(d+1)/d}$.

A check of the validity of this expression by fitting the distribution of gyration radii \autoref{eq:PofRg2} to numerical data obtained by extensive on-lattice Monte Carlo simulations at different values $N$ and $\varepsilon$ (see \autoref{sec:numerical} for details) results however in a still unsatisfactory agreement, as shown by~\autoref{fig:cmp-a4} (red curves).

We therefore introduced a new surface energy term, whose form is deduced, once again, by scaling arguments.
As shown by \citeauthor{Imbert1997}, at the coil-globule crossover, free energy is a function of the scale variable $\hat t = N^{1/2} t$~\cite{Imbert1997}.
In addition, the surface energy depends on the polymer length as $N^{(d-1)/d}$~\cite{Owczarek1993}.
In order to satisfy these two conditions, the surface energy must be of the form
\begin{equation} \label{eq:surface-energy}
    \beta F_\sigma \sim (N t^2)^{(d-1)/d}\,.
\end{equation}
By comparing the predictions of the new surface term with the data from our simulations, we find that it reproduces not only the medians and means much better (\autoref{fig:cmp-a4}, blue curves), but also all the distributions, as we will see in the next section.

The \citeauthor{Khokhlov1981}'s expression of the surface energy is similar to our \autoref{eq:surface-energy} for $d=3$, except that it does not have the same scaling variable.
It would be appropriate to test the robustness of the model for other dimensions $d \neq 3$ and to support the expression of our surface energy that predicts another dependence in $d$.

Introducing the surface energy term of~\autoref{eq:surface-energy},
the free energy of the attractive self-avoiding walk finally reads
\begin{align} \label{eq:corrected-free-energy-isaw-with-c}
    \beta F_N(t|\varepsilon) &= a_1(\varepsilon) \cdot N t + a_2(\varepsilon) \cdot N t^2 + a_3(\varepsilon) \cdot (N t)^{-q} \nonumber \\
&+ a_4(\varepsilon) \cdot (N t^2)^{2/3} +  c_N(\varepsilon)\ln N t\,.
\end{align}
with $c_N(\varepsilon)$ given by \autoref{eq:corrected-cN} and the introduction of a new function $a_4(\varepsilon)$ to be inferred from the simulations.\\

In summary, we derived an explicit formula for the free energy, \autoref{eq:corrected-free-energy-isaw-with-c},
including a surface term and a modified logarithmic term, both of which were fitted using extensive simulations.
The new free energy has the important benefit of factorizing its dependence on the polymer length $N$ and on the interaction parameter $\varepsilon$.
Thanks to this, it becomes possible to explicit its expression by determining the $\varepsilon$-dependent parameter $\theta(\varepsilon) = (a_1, a_2, a_3, a_4, c_0, c_1, c_2)$ on a numerical basis, as shown in the next section.

\section{Validation of the free energy formula on simulated \texorpdfstring{$R^2$}{R2} distributions}
\label{sec:numerical}

\subsection{Simulation of Interacting Self-Avoiding Walks (ISAW)}
\label{sec:simulations}

\begin{figure*}[ht!]
    \includegraphics[width=.8\textwidth]{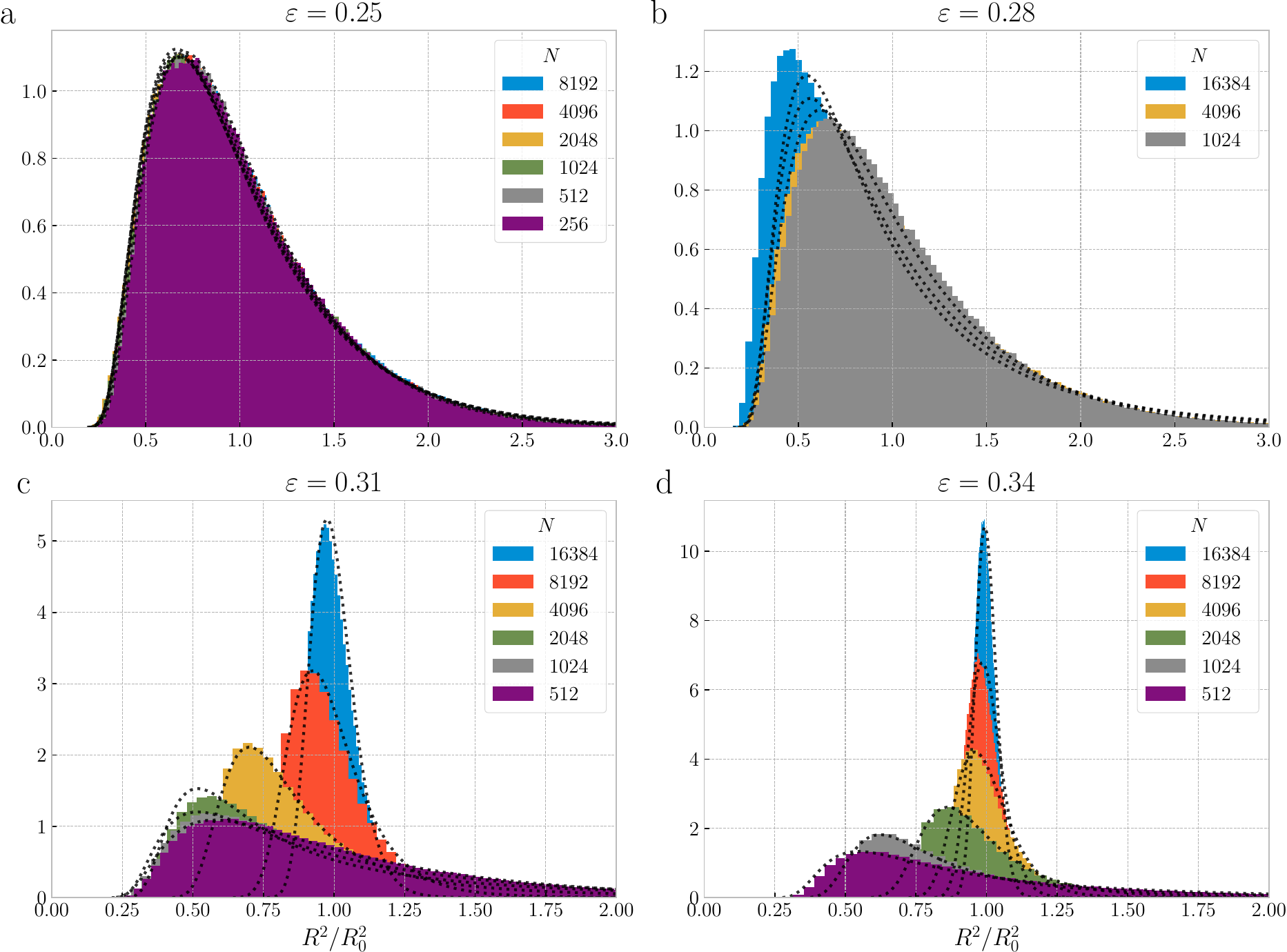}
    \caption{
        Comparison of the agreement between Monte Carlo simulations (histograms) and theory (dotted lines) for different lengths $N$ at different values of $\varepsilon$ (\textbf{a}, \textbf{b}, \textbf{c}, \textbf{d}).
        The distributions are normalized by $R_0^2$, the expectation of the radius of gyration squared.
    }
    \label{fig:data-histograms}
\end{figure*} 

While off-lattice polymer models are more realistic, and despite some recent advances in defining faster algorithms~\cite{Schnabel2022}, on-lattice models remain the most studied~\cite{Grassberger1995,Grassberger1997,Vogel2007}, as they allow simulating polymers of much higher lengths.
We therefore chose to simulate interacting self-avoiding walks (ISAW) on a cubic lattice~\cite{Tesi1996}.
The conformations of a polymer of $N$ monomers are sampled thanks to the Metropolis algorithm with reptation moves~\cite{Wall1975}, much more effective in exploring the configuration space for systems of high monomer densities (as globules), than \eg{} the pivot algorithm\cite{Lal1969}.
The simulated lattice contains a unique chain, in an infinite volume.
Effective interaction between monomers, which depends on the balance of free energy between monomer-monomer and monomer-solvent interaction, are simply modelled by introducing an energy gain when two non-contiguous monomers are in contact, \ie{} on neighboring lattice sites.
From the number $m$ of non-bonded nearest neighbor contacts we thus get the energy of a given conformation $-mJ$, hence the dimensionless energy $m \varepsilon = mJ/k_BT$.

We performed simulations first for $42$ values of $N$ varying between $2^3$ and $2^{13.25}\simeq 10\, 000$ (in logarithmic scale) and $28$ values of $\varepsilon$ in the range $0.00$ to $0.27$;
second for $48$ values of $N$ varying between $2^3$ and $2^{14.5}\simeq 23\, 000$ (in logarithmic scale) and $12$ values of $\varepsilon$ in the range $0.28$ to $0.39$.
For each pair $(N,\varepsilon)$, $2^{19} \simeq 524\, 000$ chains are simulated and the corresponding instant squared radius of gyration $R^2$ (see \autoref{eq:RG1}) calculated, this leading to a dataset $\{ R^2_i \}_{N,\varepsilon}$.
For each condition, \ie{} for each pair of parameters ($N, \varepsilon$), the simulation started with a fully extended conformation of the polymer.
We verified that the polymer relaxed toward equilibrium, with a relaxation timescale $\tau_R$ that scales as $N^2$.
We chose to define $\tau_R=0.5\, N^2$ in units of \emph{accepted} Metropolis steps and fixed an equilibration period of $2^7 \, \tau_R$, over which the interaction parameter $\varepsilon$ was progressively changed until the target value of the system was reached.
Note that, if $\tau_R$ does not depend explicitly on $\varepsilon$, the effective equilibration time does increase with $\varepsilon$, as there are more rejected steps when the interaction becomes stronger.
Once the system equilibrated, the gyration radius was computed every $2^{-5} \, \tau_R$ step over $2^{14} \, \tau_R$ steps.

\begin{figure*}[ht!]
    \includegraphics[width=.8\textwidth]{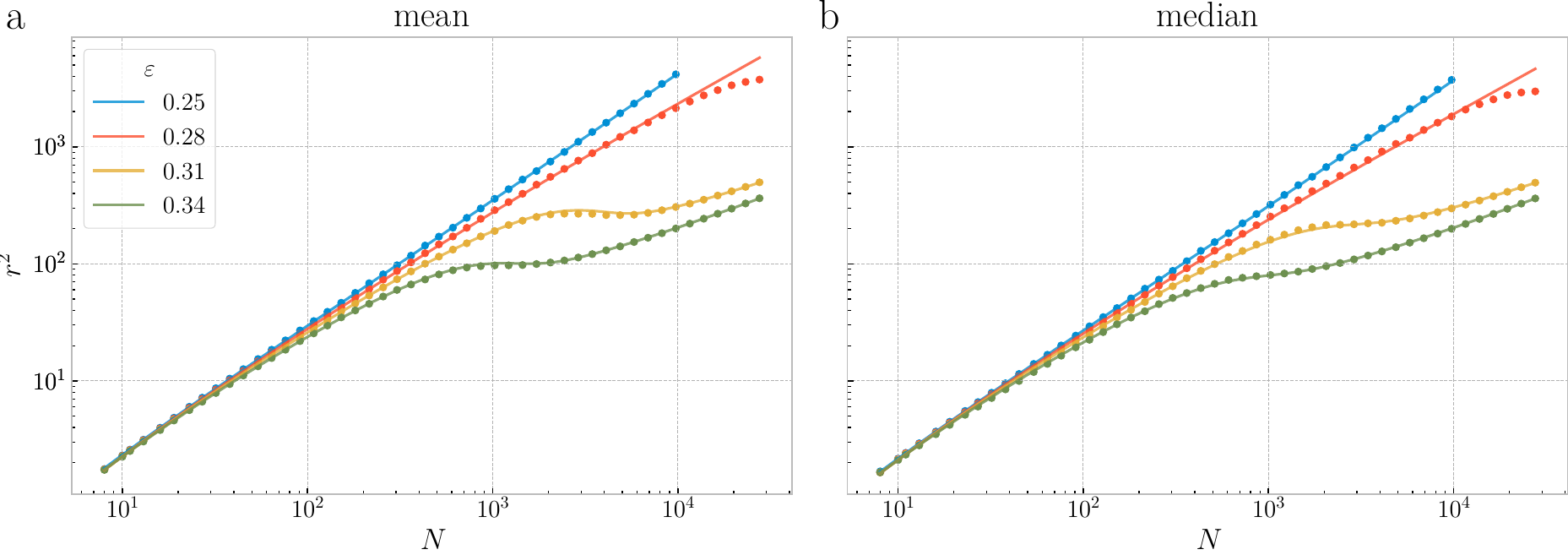}
    \caption{
        Comparison of the agreement between Monte Carlo simulations (dots) and theory (lines) both in mean (\textbf{a}) and in median (\textbf{b}) for different values of $\varepsilon$.
    }
    \label{fig:data-mean-median}
\end{figure*}

\subsection{Bayesian inference}
\label{si:bayesian-inference}

Our theoretical expression of the polymer free energy explicitly specify its dependence on both the polymer length $N$, the gyration radius $R^2$ (or equivalently the reduced density) and the interaction parameter $\varepsilon$, so to factorize these dependencies.
Our final formula includes 
seven 
parameters that need to be inferred in order to fit the data.
These parameters, that we grouped as $\theta(\varepsilon) = (a_1, a_2, a_3, a_4, c_0, c_1, c_2
)$, are expected to be functions of $\varepsilon$ only.
However, the analytic form of the functions $a_i(\varepsilon)$ and $c_i
(\varepsilon)$ are \apriori{} not known.
In order to obtain an estimation, we collected all $\{ R^2_{i} \}_{N,\varepsilon}$ data for different $N$ in a single $\{ R^2_{i,N} \}_{\varepsilon}$ dataset, then inferred $\theta(\varepsilon)$ from the dataset, for each $\varepsilon$ through Bayesian inference.
We build the probability density function for $R^2$, ${\cal P} (R^2 \big\vert \varepsilon, N)$, using the explicit free energy  \autoref{eq:corrected-free-energy-isaw-with-c} 
and maximizing the log-likelihood function
\begin{equation}
    \mathscr{L}( \theta(\varepsilon)) \equiv \ln \prod_{i,N} {\cal P} (R^2_{i,N} \big\vert \varepsilon, N) 
\end{equation}
\noindent
according to the Bayesian inference approach.
We used a uniform \emph{prior} probability distribution (\emph{naive} Bayesian inference), by fixing only the positiveness of the fitting parameters.
We then performed Goodman \& Weare's Affine Invariant Markov chain Monte Carlo Ensemble sampler~\cite{ForemanMackey2013} to infer the \emph{posterior} distribution for the parameters $ \theta(\varepsilon)$.

\subsection{Energy dependence of the free parameters}
\label{sec:Energy dependence of the parameters}

We proceeded with the fitting of the $\varepsilon$-dependence of the seven 
coefficients gathered in the vector $\theta(\varepsilon)$.
Examples of dataset distributions for four different values of $\varepsilon$ (and different lengths $N$) are represented as histograms in \autoref{fig:data-histograms}.
Note the x-axes ranges have been adapted to the data extent.
Similar histograms are obtained for each pair of parameters ($N, \varepsilon$).

From these data we extracted the mean and median values of the radius of gyration, which are given in \autoref{fig:data-mean-median} as functions of $N$ for the same four values of $\varepsilon$ (dots).
The most striking feature of \autoref{fig:data-mean-median} is the presence of an inflection point for $\varepsilon > \varepsilon_\Theta$.
Moreover, the radius of gyration may even become non-monotonous with $N$ (see \autoref{fig:data-mean-median} for $\varepsilon=0.31$).
This behavior is quite unusual among critical phenomena and leads to dramatic finite-size effects.
This intermediate region where the radius of gyration varies more slowly,  and thus differs from both the scaling behavior of coil and globule (with potential difficult interpretation of experimental data, see \cite{Lesage2019} for an example), defines the crossover region~\cite{Grassberger1995}.

\begin{figure*}[ht!]
    \includegraphics[scale=0.35]{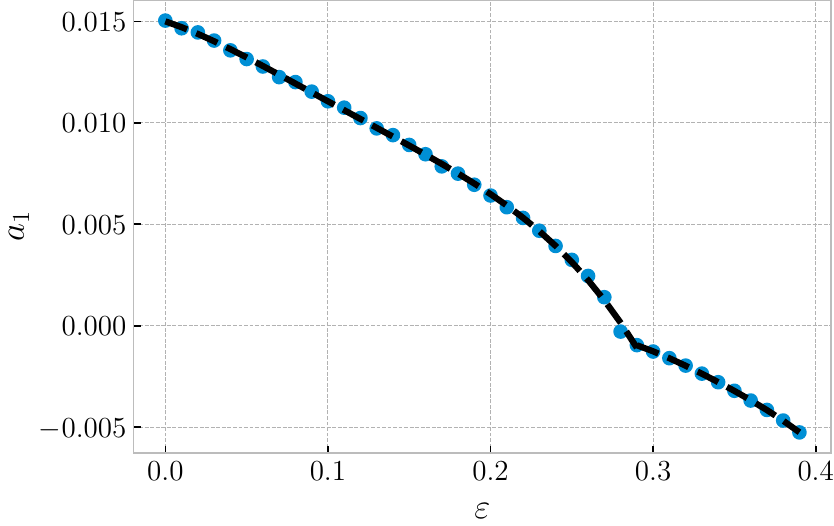}
    \includegraphics[scale=0.35]{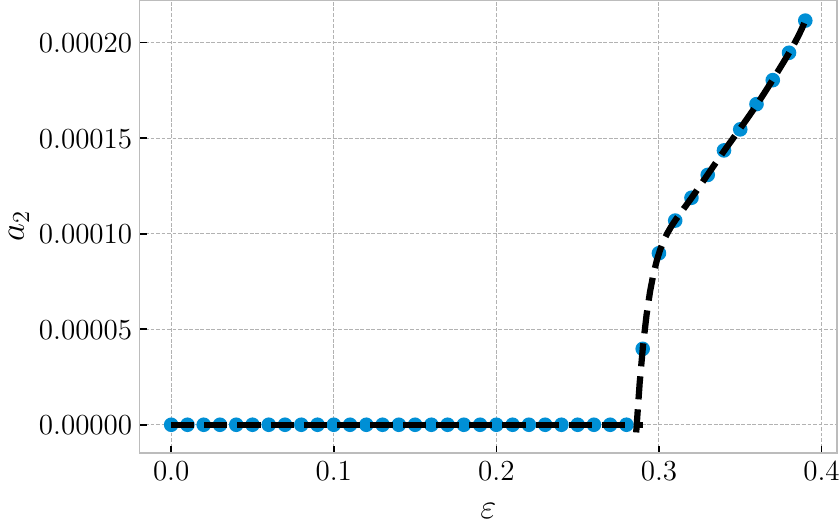}
    \includegraphics[scale=0.35]{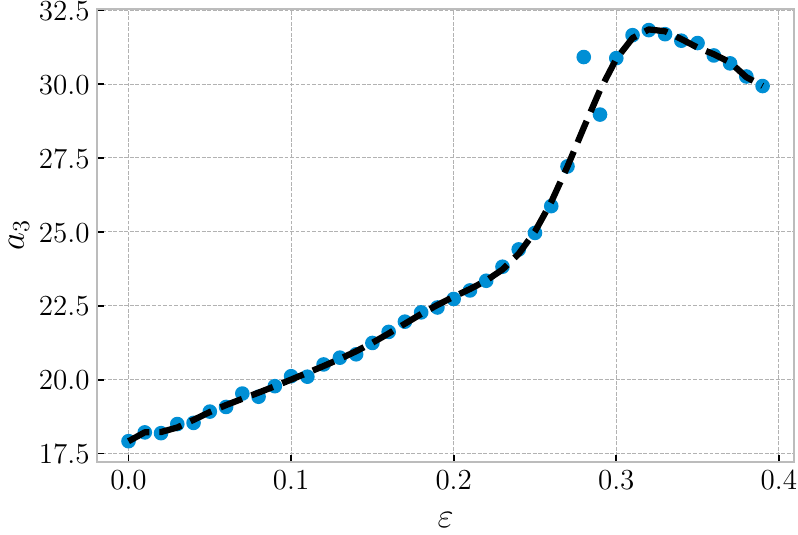}
    \includegraphics[scale=0.35]{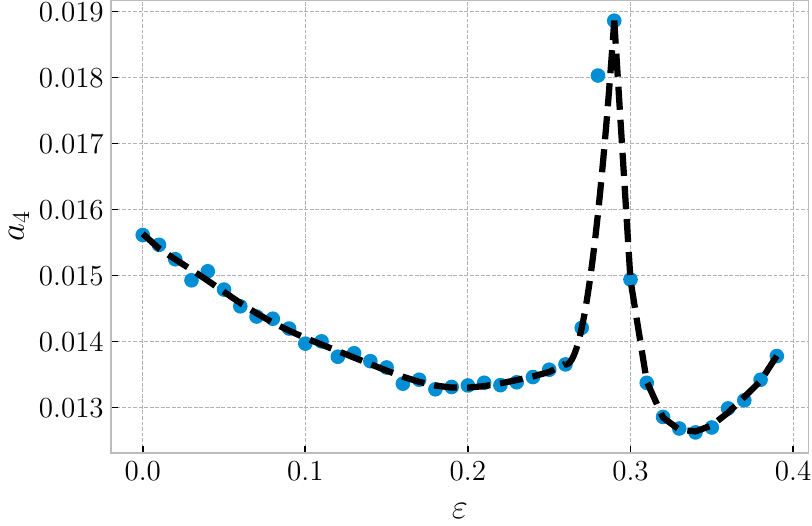}
    \includegraphics[scale=0.35]{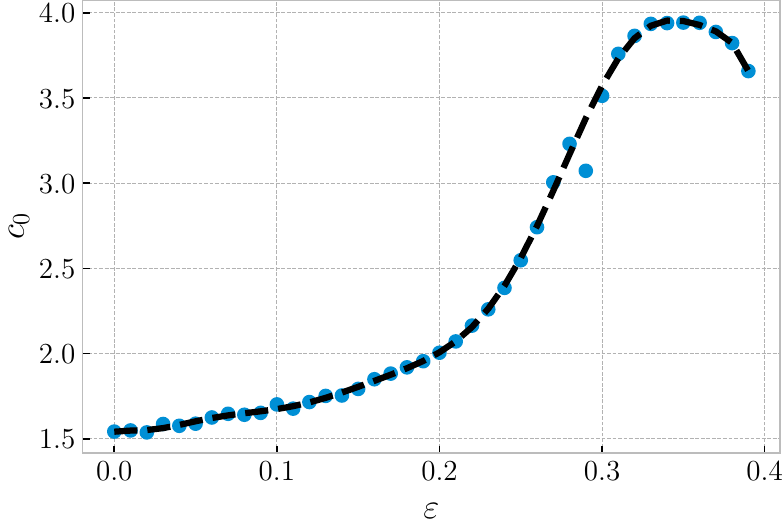}
    \includegraphics[scale=0.35]{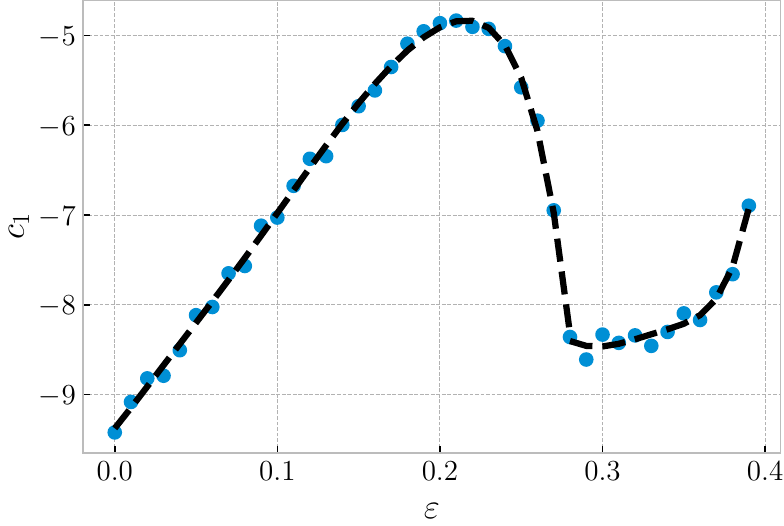}
    \includegraphics[scale=0.35]{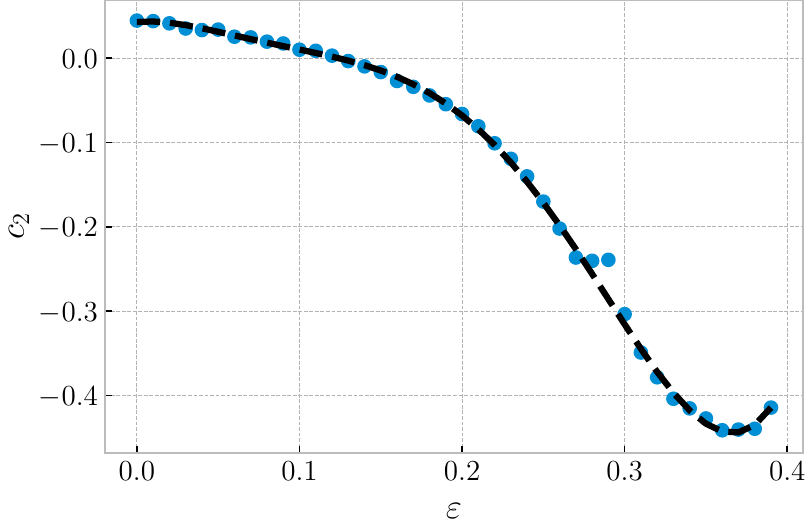}
    
    \caption{Inferred evolution of the free energy parameters $a_1,a_2,a_3, a_4$ and $c_0, c_1,c_2$
    with respect to $\varepsilon$. The dots are the Bayesian fit results, the dashed lines correspond to interpolation functions. These interpolation functions can be reproduced by using the code freely available at 
    \url{https://github.com/antonylsg/model-jupyter.git}.    }
    \label{fig:parameters}
\end{figure*}

We used the Bayesian procedure described in the methods section and applied it to each different value of the energy parameter $\varepsilon$.
We were able to obtain the best estimates for each of the parameters $a_i(\varepsilon)$, $c_i
(\varepsilon)$.
The results are shown in \autoref{fig:parameters}.
The obtained values for $a_i(\varepsilon)$ and $c_i
(\varepsilon)$ are compatible with regular functions of $\varepsilon$, with a remarkable change in behavior near a critical value $\varepsilon_c \simeq 0.3$.
We performed a polynomial regression to obtain an interpolation of the functions $a_i(\varepsilon)$ and $c_i
(\varepsilon)$.
This procedure is sufficient to reproduce the curves with good accuracy (see \autoref{fig:parameters}, dashed lines).
 
It is interesting to recall that the prefactor $a_1(\varepsilon)$ is comparable to the second virial coefficient and drives the coil-globule phase transition.
$a_1(\varepsilon)$ vanishes when the attractive interactions counterbalance the effective influence of excluded volume, at the $\Theta$-point~\cite{Wang2017}.
When $a_1(\varepsilon)$ is \emph{positive}, excluded volume interactions dominate and the resulting distribution of conformations is that of \emph{coil states}, while when it is \emph{negative}, attractive interactions dominate and the resulting distribution of conformations is that of \emph{globule states}.
In agreement with these predictions, our fit of $a_1(\varepsilon)$ gives $\varepsilon_\Theta \simeq 0.28$, consistently with previous estimates
\cite{Grassberger1995,Vogel2007}.

The coefficient $a_2(\varepsilon)$ is positive.
This result was expected: $a_2(\varepsilon)$ must be strictly positive so that the probability density function ${\cal P} (R^2 \big\vert \varepsilon, N)$ remains integrable.

The $a_4$ term isn't relevant for self-avoiding walks ($\varepsilon=0$), as there is no energy penalty for a monomer to be on the surface.
In fact, we reproduce the probability distribution very well without it.
In addition, it contributes to the free energy only for the densest ($t \gg 1$) and thus globular conformations, because of the form $N t^2$.

Interestingly, some of the parameters display a discontinuous derivative close to $\varepsilon_\Theta$.
For these parameters, the regression has been performed separately over two ranges of $\varepsilon$, except for $a_4$ which required three ranges.

Although the different parameters are globally justified by different contributions to the free energy, we notice some anomalies in their behavior as determined by the Bayesian regression, and in particular the fact that $a_4$ increases for $\varepsilon \to 0$.
This is unexpected since the surface term should be negligible for coils.
Such an effect is probably a consequence of an overfitting in these conditions (pure coil).
One should therefore keep in mind that, while the free energy \autoref{eq:corrected-free-energy-isaw-with-c} 
accurately describes the coil-globule transition and reproduces correctly the histograms of the gyration radius, compensation effects between parameters may exist in regions further away from the transition.

\subsection{Theoretical prediction compared to simulated histograms and averages}
\label{sec:performance}

We assess \aposteriori{} the quality of the free energy expression and of the resulting parameters $\theta(\varepsilon) = (a_1, a_2, a_3, a_4, c_0, c_1, c_2)$ by comparing the theoretical predictions - obtained from \eqref{eq:PofRg2} and \eqref{eq:corrected-free-energy-isaw-with-c} using the Bayesian estimates of $\theta(\varepsilon)$ - to the simulated distributions of $R^2$. The full distributions, shown in \autoref{fig:data-histograms}, provide the most stringent test. A more synthetic view is given in \autoref{fig:data-mean-median}, which plots the mean and median values of $R$ as a function of $N$ for several $\varepsilon$, in the same spirit as \autoref{fig:cmp-a4}; it also makes explicit that, for sufficiently large $N$, only two scaling regimes remain, corresponding to coil ($\varepsilon < \varepsilon_\Theta$) and globule ($\varepsilon > \varepsilon_\Theta$) conformations.

Both figures show that our approach reproduces the histograms ${\cal P} (R^2 \big\vert \varepsilon, N)$ - and \emph{a fortiori} their mean and median - for essentially all $N$ and $\varepsilon$. The only notable deviations appear very close to the critical value of $\varepsilon$, as illustrated in \autoref{fig:data-mean-median} for $\varepsilon=0.28$: the corresponding inflection point lies beyond the explored range of $N$, so the simulated histograms are only approximately fitted by the theoretical curves.
The origin of this discrepancy 
may therefore be numerical rather than theoretical: obtaining sufficiently representative samples of the globule phase becomes practically overwhelming as we approach the critical point.
Interestingly, however, the scaling variable introduced in \autoref{eq:with_hat_t} is equal to
\begin{equation}
    \hat{t} = N^{\frac{1}{2}}t = N^{\frac{1}{2}}\Bigl({\frac{N}{R^d}}\Bigr)^{\frac{1}{\nu d-1}}
= \Bigl({\frac{N^{\nu_\theta}}{R}}\Bigr)^{\frac{d}{\nu d-1}}
\end{equation}
with
 $\nu_\theta = {(\nu + {1}/{d})}/{2}$.
This means that the polymer at the $\Theta$-point would be scale invariant with a fractal dimension $d_f = 1/\nu_\theta$, which would also make $\nu_\theta$ precisely the Flory exponent at the $\Theta$-point. As $d=3$, this would give $\nu_\theta = (\nu+1/3)/2 \simeq 0.46$, the analogue for an interacting self-avoiding walk of the classical exponent $\nu_\theta=1/2$ for an ideal chain at the $\Theta$-transition.
However, this value assumes that the dominant term of the free energy remains quadratic in $t$ (\ie{} that the exponent $n$ of the $a_2$ term introduced in \autoref{eq:free-energy-expansion-of-t-around-theta} is exactly $2$) all the way down to the $\Theta$-point itself. Since $d=3$ is precisely the upper critical dimension of the $\Theta$-transition, where mean-field behavior is expected to become asymptotically exact (up to logarithmic corrections), consistency instead requires $\nu_\theta=1/2$. A treatment consistent with this constraint would fix $\nu_\theta=1/2$ to its established value and derive, rather than assume, the effective exponent $n$, which would give $n\simeq1.53$ instead of $2$. This point could not be tested against the numerical data available for the present work, and is left for future investigation.

Overall, the very accurate agreement between simulations and theoretical expressions gives the best evidence that the proposed model describes correctly the behavior of on-lattice polymers, and allows reproducing in particular the finite-size effects previously discussed.
Furthermore, the good agreement with the whole $R$ distribution confirms the choice of the different terms in the free energy.

\begin{figure*}[ht!]
    \includegraphics[width=.8\textwidth]{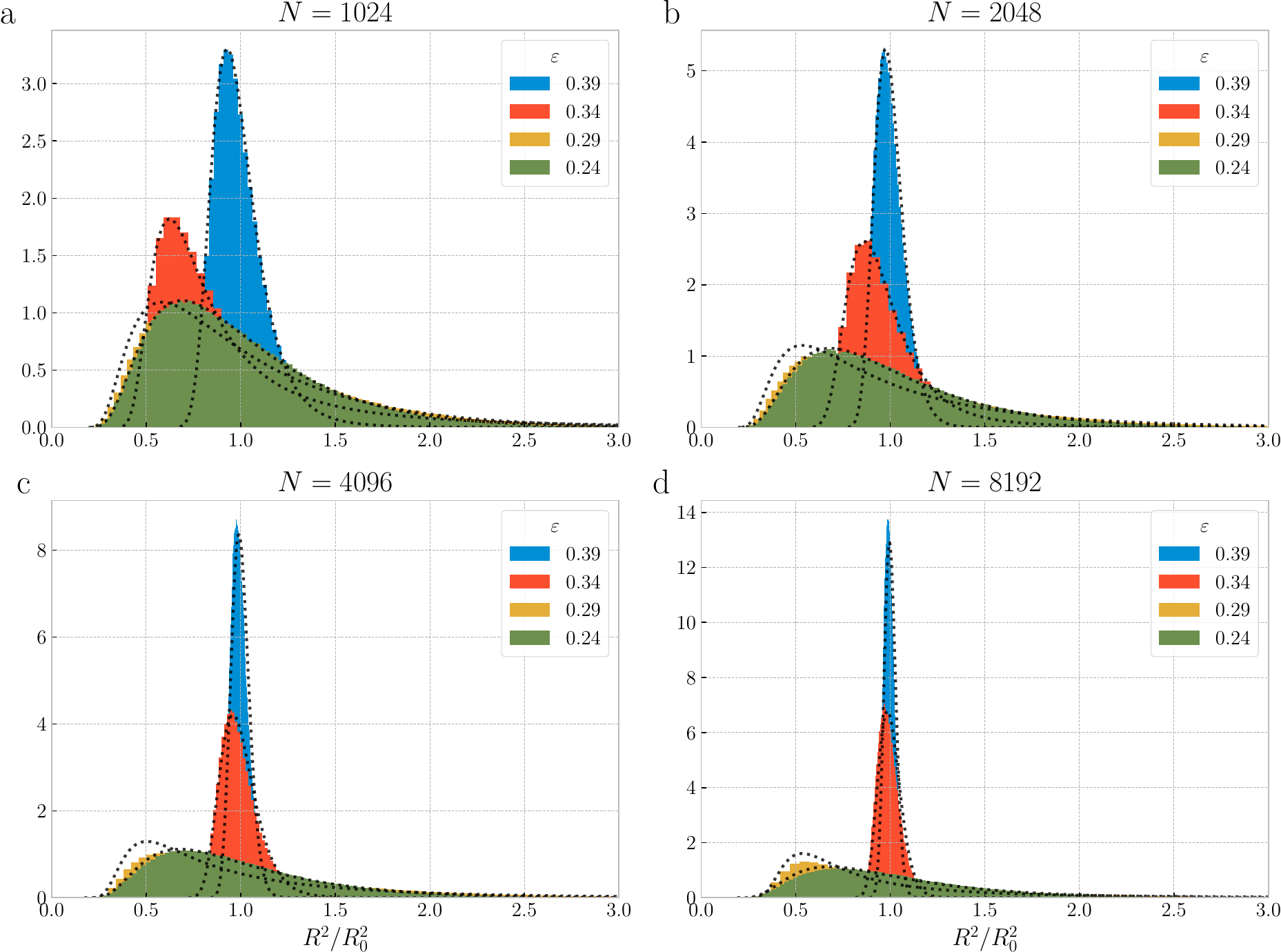}
    \caption{Monte Carlo simulations (histograms) and the corresponding theoretical curves (dotted lines) as for \autoref{fig:data-histograms} but this time for different $\varepsilon$ at different values of the lengths $N$.
        The distributions are normalized by $R_0^2$, the expectation of the radius of gyration squared.}
    \label{fig:histo-eps}
\end{figure*}

The analysis of the distribution of the radius of gyration - as a direct representation of the free energy - is, moreover, particularly informative about the phase transition process.
In \autoref{fig:histo-eps}, we show it for different values of $\varepsilon$ for given $N$, and this for four different values of $N$.
This is simply an alternative view to the \autoref{fig:data-histograms}, but it allows to follow the evolution of the radius of gyration distribution with an increasing $\varepsilon$, \ie{} as a function of the solvent quality.
The resulting distributions exclude the existence of coexisting phases at the transition, and support a continuous transition from a swollen to a dense phase when increasing $\varepsilon$ in agreement with the standard theory ~\cite{Grosberg1994}.

\subsection{Application: Phase diagram}

\begin{figure*}[ht!]
    \includegraphics[width=.65\textwidth]{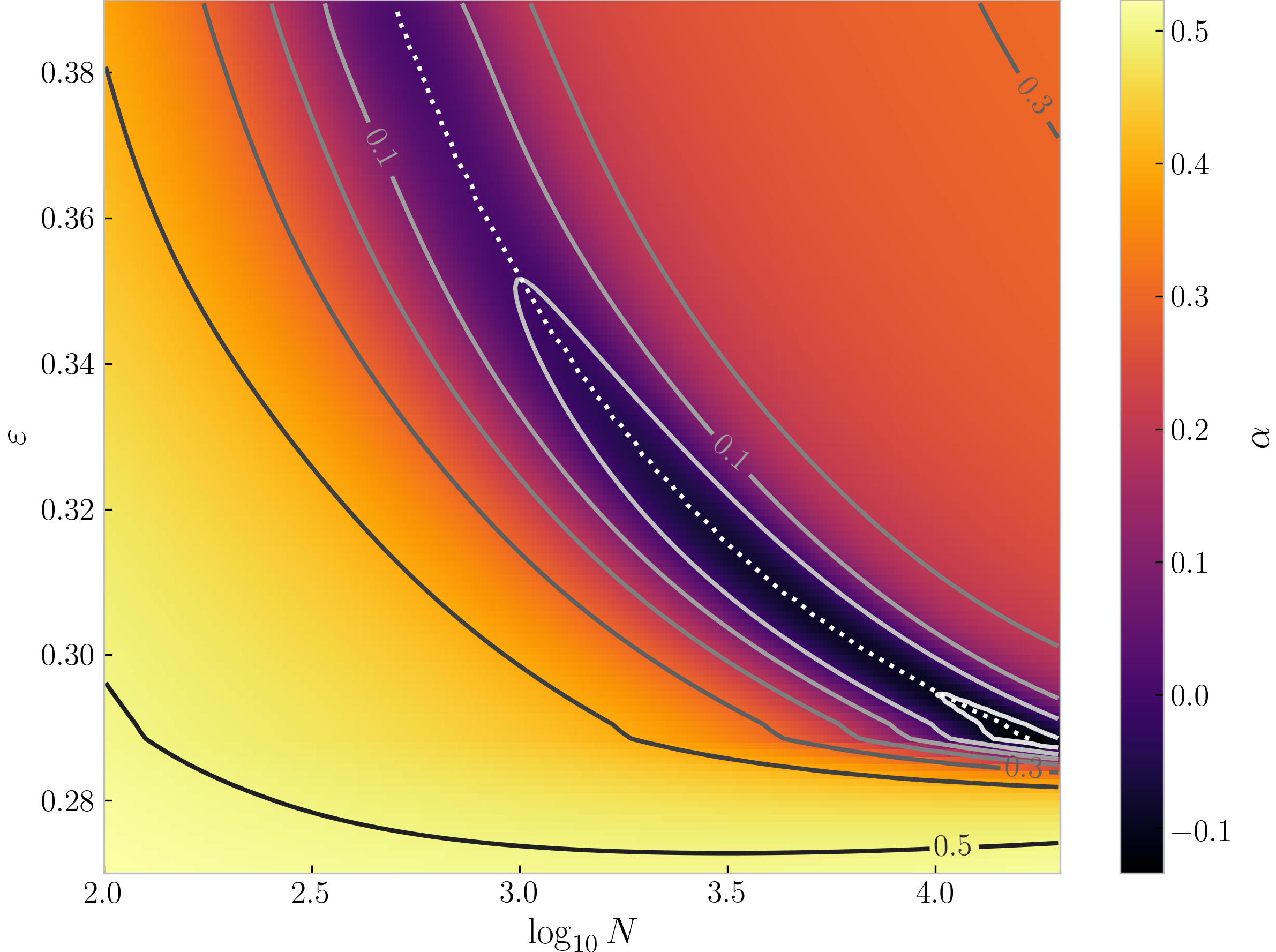}
    \caption{
        Coil-globule phase diagram.
        The colorscale corresponds to the derivative of the function $\bar R(N)$ for a given $\varepsilon$, here displayed as a function of $\varepsilon$ and $\log_{10}N$.
        The white dotted line is the line of inflection points of $\bar R(N)$, \ie{} the line of minima of $d\bar R(N)/dN$, for given $\varepsilon$ in log-log scale.
        This line can be fitted indifferently by: $\varepsilon_N = a N^b + \varepsilon_\Theta$, with $a=3.31 \pm 0.06$, $b=-0.538 \pm 0.003$, $\varepsilon_\Theta = 0.2719 \pm 0.0003$; or $\varepsilon_N = a N^{-1/2} + b N^{-1} + \varepsilon_\Theta$, with $a=2.44 \pm 0.02$, $b=4.6 \pm 0.3$, $\varepsilon_\Theta = 0.2703 \pm 0.0002$.
    }
    \label{fig:phase-diagram}
\end{figure*}

Finite size effects enrich the features of the coil to globule transition, and represent, for this reason, an excellent reference benchmark to test the efficiency of a model.
As an application of our formula, we have therefore aimed at reproducing a phase diagram of the coil-globe transition based on the theoretical expression, with the idea of comparing it with pre-existing theoretical and numerical results.
In the thermodynamic limit, the only physical parameter is the energy parameter $\varepsilon$, the phase diagram is one-dimensional and shows a unique critical point, named $\Theta$-point, at $\varepsilon_\Theta$.
This is a second order phase transition, which means that, at the critical point, the specific heat of the system diverges~\cite{Lifshitz1978}.
In the finite $N$ case, however, two physical parameters, $\varepsilon$ and $N$, determine the state of the system.
Therefore, the phase diagram is two-dimensional.
As we noted before, the distribution of microscopic states of polymers smoothly changes from one type that can be identified with a coil phase to another type that can be identified with a globule phase.
As both the sharpness and the position of the phase transition depend on $N$, the identification of the $N$-dependent transition point $\varepsilon_N$ is rather fuzzy.

A first theoretical prediction for the scaling of $\varepsilon_N - \varepsilon_\Theta$ with $N$ has been proposed by  \citeauthor{Lifshitz1978}~\cite{Lifshitz1978} and reads
\begin{equation}
\label{eq:thetan}
    \varepsilon_N = A \,{N}^{-1/2} + \varepsilon_\Theta\,,
\end{equation}
where $A$ is a constant.

Numerical estimations are based either on the vanishing of the second virial coefficient~\cite{Grassberger1995}, or on the condition of divergent specific heat~\cite{Vogel2007,Rampf2006}, or on a specific scaling of $\langle R^2\rangle$ with $N$~\cite{Rampf2006}.
These derivations introduced correction terms of different types.
In Ref.~\cite{Vogel2007}, \citeauthor{Vogel2007} performed a comparison of these different expressions and showed with on-lattice simulations that the boundary between the two phases is best fitted by a function of the form
\begin{equation}
    \label{eq:thetanVogel}
    \varepsilon_N = A N^{-1/2} + B N^{-1} +  \varepsilon_\Theta.
\end{equation}

Starting from our previous results, it seems therefore appealing to dress a phase diagram for the coil-globule transition based on the behavior of $\bar R$.
A simple idea is to focus on the local slope of the $N$-dependent curves $\bar R(N)$ for given $\varepsilon$, as an alternative order parameter.
Indeed, this slope decreases in the crossover region and we can expect it to reach a minimum close to the phase transition.
The colormap of \autoref{fig:phase-diagram} shows the values of the slope $d\bar R(N)/dN$ in the ($N$, $\varepsilon$) plane.
The decreasing of $d\bar R(N)/dN$ along the transition line is clearly visible.

We checked the behavior of the inflection point (slope minimum) and found that it can be rather well approximated by the function \autoref{eq:thetanVogel}, with $\varepsilon_\Theta = 0.2703 \pm 0.0002$, which is compatible with previous estimates of this asymptotic limit.
Nevertheless, the data are also compatible with a power low behavior $a N^{b} + \varepsilon_\Theta$, with $a=3.31 \pm 0.06$, $b=-0.538 \pm 0.003$ and $\varepsilon_\Theta = 0.2719 \pm 0.0003$ (see \autoref{fig:phase-diagram}).
Instead, the theoretical behavior in ${N}^{-1/2}$, expected from \autoref{eq:thetan} for the transition, poorly correlates with the inflection point (data not shown).

It should be noticed that the $\Theta$-point does not generally correspond to the transition point, but rather to conformations that are still in the coil regime~\cite{Foldes2021}, although close to the transition.
The $\Theta$-point only coincides with the transition point \emph{in infinite-size}.
Interestingly, a rather different approach to the study of the finite-size coil-globule transition line $\varepsilon_\Theta(N)$ highlighted similar discrepancies between different ways of defining the critical transition point~\cite{Foldes2021}.
Here, we introduced an independent definition of the transition line based on the infection point of $\bar R(N)$ and quantitatively recover the typical critical behavior.

\section{Conclusions}
\label{Conclusions}

Using extensive Monte Carlo simulations for a wide range of solvent conditions together with scaling arguments, we were able to derive an explicit formula of the free energy of a chain of $N$ interacting monomers as a function of its gyration radius, which works remarkably well at any solvent quality up to the immediate vicinity of the coil-globule transition (for large $N$, \ie{} $\varepsilon$ between $0.27$ and $0.30$), where the formula is slightly less accurate.
To come to such an agreement, we could not use less than 
seven 
parameters in the free energy, that vary as a function of the interaction energy but do not depend on $N$.
This is, in fact, a relatively small number of parameters if we consider that it enables to fit not only the variation of the mean radius of gyration but of the whole distribution of the instant radius of gyration for every length $N$ and interaction energy $\varepsilon$.
We emphasize that these distributions can take very different shapes as the polymer phase diagram is explored, yet our formula accurately fits all these distributions with $N$-independent parameters.
This is a significant improvement with respect to the \citeauthor{Sanchez1979}'s mean-field  theory~\cite{Sanchez1979} which is based on a Boltzmann-weighted Flory-Fisk distribution of the instant radius of gyration~\cite{Hofmann2012}.
Indeed, the Flory-Fisk distribution, even though it seems to reproduce reasonably well the ratios of the first moments of the exact distribution, is a poor fit of the simulation results~\cite{Vettorel2010}.

We emphasize again that our formula provides the explicit dependence of the free energy on $R$ and $N$, and that the fitting parameters contain, by themselves, the whole energy dependence of the system behavior.
Moreover, it is interesting to note that the regularity of the $\varepsilon$-dependent parameter fitting curves (\autoref{fig:parameters}) suggests further rationalization of the system behavior.
We have already shown how some of the parameters have a clear interpretation and/or are expected to vanish in specific phases.
Our present results open the prospect of a model where the heuristic component would be further reduced, which will be explored in future works.

Another advantage of the explicit free energy formula introduced in this work is that it can be used to map more complex off-lattice~\cite{Parsons2006} models, including specific coarse-grained polymer models~\cite{Rampf2006,Vettorel2010}, on a on-lattice ISAW.
This is not straightforward.
Indeed, a precise mapping between the energy per contact on the lattice and the parameters of the off-lattice force field is necessary to obtain a clear correspondence between equivalent states.
Preliminary tests based on spectral analysis~\cite{Foldes2021} show that the large-scale characteristics of an off-lattice bead-spring polymer model with Lennard-Jones interactions are qualitatively the same as for the lattice-based model studied in this work (data not shown).
However, a quantitative comparison, especially concerning the transition line and the asymptotic theta point, must be made, and may be the subject of future work.

Finally, as a general conclusion, we wish to discuss some practical motivations of our work.
In physical chemistry, recent developments of the quasi-elastic neutron scattering techniques, and more particularly of methods based on Neutron Spin Echo, can lead to the fluctuations of the shape of scattering molecules \cite{Wu2021}, including the radius of gyration of polymer chains.
Modern applications in biology are also primarily concerned by the description of such an observable.
This is indeed particularly relevant to Intrinsically Disordered Proteins (IDPs) for which the quantification of the solvent quality is of foremost importance~\cite{Soranno2020}.
Moreover, the 3D genome organization inside the cell nucleus is one of the most challenging questions of modern cell biology.
Significant advances have been achieved in the characterization of the complex, multiscale polymer that contains the genetic material, known as \emph{chromatin}.
The spatial arrangement of chromatin appears to play a key role in gene regulation, especially at the crucial scale of \textit{epigenetic domains} \cite{Cortini2016}, whose configurations are well described by polymers close to the coil-globule transition \cite{Lesage2019}.
Obtaining a physical description of the spatial organization of chromatin at the level of  epigenetic domains is therefore a crucial issue.
Experimentally, this goal is challenging, but the resolution limitation of traditional optical imaging has recently been overcome by super-resolution~\cite{Boettiger2016, Cattoni2017},  notably confirming a clear distinction between regions of chromatin that appear to be swollen and regions with a more dense, globular conformation.
In such experiments, the radius of gyration is essentially the only measurable order parameter, and the ability to fit its distribution instead of just its mean value potentially offers a much richer exploitation of the data.
Therefore, there is a need for new methodologies to extract information from experimental distributions of the radius of gyration.
The explicit expression of these distributions here derived from the system free energy meets precisely this need.

\begin{acknowledgments}
The authors would like to acknowledge networking support by the EUTOPIA COST Action CA17139.
This work has been partially supported by the ANR project ANR-19-CE45-0016.
\end{acknowledgments}

\bibliography{references-Lesage}

\end{document}